\documentstyle[epsf,prl,twocolumn,aps]{revtex}
\begin{document}
\title{ Persistent Hall 
Voltage and Current in the Fractional Quantum Hall Regime}
\author{ Stefan Kettemann }
\address{ {\it 
Weizmann Institute of Science
Department of Condensed Matter Physics, Rehovot 76100, Israel \\
and Max- Planck Institute f\" ur Physik
 Komplexer Systeme, Au\ss enstelle Stuttgart, Heisenbergstr.1,
 70569 Stuttgart, Germany}}
\date{\today }
\maketitle
\begin{abstract}
  The persistent Hall voltage and current in an isolated  annulus
 in a strong perpendicular magnetic field,
 at filling factor $\nu =1/q$, and  
 in the presence of a weak constriction 
  is obtained as a function of temperature, and flux piercing
 the annulus.  
 A thermodynamic Hall conductance is found which has a universal
value even with back scattering at the constriction.  
\end{abstract}
\section{Introduction}
The trial wavefunction proposed by 
Laughlin was established to 
describe very well the ground state of a two- dimensional
electron gas 
in a strong perpendicular magnetic field 
at filling factors $\nu=1/m$, m odd
 as an incompressible state
\cite{laughlin83}.
However, it is not yet clear if this implies that
the excitations of this ground state have fractional
charge and fractional statistics.

In this article we 
consider  
 a 2- dimensional  annulus  
 in a strong magnetic field
 which condenses the electrons to a fractional quantum hall state.
 We will call this a FQH- annulus in the following.
The Laughlin wave function for a pure FQH- annulus   
 when the filling factor
satisfies $\nu=1/m$,
 where $m$ is an odd integer,
is given by\cite{5}: 
\begin{equation}
\psi = \prod_{i=1}^N \mid z_i \mid^{\varphi} z_i^n
\exp ( -\frac{\mid z_i \mid^2}{4 l_B^2} ) \prod_{j=1}^{i-1}
(z_i - z_j)^{1/\nu},
\end{equation}
where $n+\varphi$ is the total flux piercing the FQH- annulus.
The inner and outer radii of the incompressible FQH liquid in
the annulus are given by $ r^2=2 l_B^2 (n+ \varphi)$ and
$R^2=2 l_B^2(n + N + \varphi)$, respectively, N being the number of
fluxquanta in the area of the annulus.
 Increasing the flux $\phi$ 
through the annulus one thereby increases the radii of the ring.
Since the annulus is confined by a certain edge potential,
this decreases the energy of the highest filled state at
 the inner 
edge while the one at the outer edge is increased.
 Increasing the flux further, until one flux quantum $\phi_0=hc/e$
 more  pierces the annulus,
 the ground state  of the
FQH- annulus  is identical  to the one without flux. However,
in order to relax to this ground state, equilibration
between the edges of the FQH- annulus is necessary.

 The thermal equilibration can  be due to backscattering across
the bulk, since we consider an isolated  FQH- annulus
which is not connected to any leads. 
 It was predicted by Thouless and Gefen, that 
thermodynamic properties of such a system have flux periodicity $\phi_0$
due to the existence of  families of 
states of the FQH- annulus, which are connected by a physical mechanism,
the finite back scattering amplitude
of fractionally charged quasiparticles.
 If only electrons could backscatter from edge to edge,
the flux periodicity would be enhanced,
in violation of the Byers and Young theorem\cite{bye}.
 Gefen and Thouless argued 
that this could be  a proof of the existence of fractional
charge excitations.

\begin{figure*}[bhp]\label{fig1}
\centerline{\hbox{\epsfbox{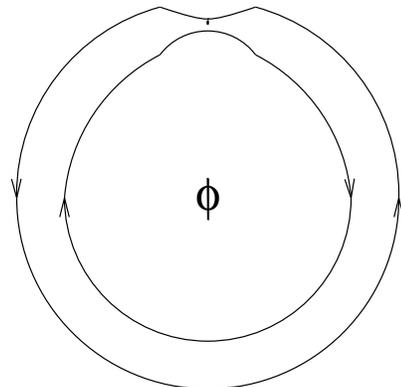}}}
\caption{ An annulus of a 2- dim. Electron- gas with a weak constriction,
 in a strong magnetic field, the arrows denote the chirality of the
 edge states}
\end{figure*}

Thouless and Gefen  considered
in Ref. \onlinecite{5} a nonequilibrium situation in the
presence of leads, with a time dependent flux through the annulus.
 Here, rather we want to
restrict us  to the thermal equilibrium of an 
isolated annulus in the presence of a weak constriction,
 see Fig. \ref{fig1}.
The other limit of a strong constriction,
 the
  weak link limit of the FQH- annulus 
 is considered in a separate article\cite{prl}.

   We will use a  model of the FQH- annulus
to establish not only the periodicity of the persistent  current,
but also its magnitude.
The Fractional Quantum Hall liquid
has gapless excitations 
at its edges while the bulk excitations have a gap\cite{wen}.
 In this article we restrict ourselves to the case of a sharp edge potential.
Then,  the FQH-ring has two branches of edge excitations, one at the 
inner and one at the outer edge.
 Special care has to be taken of the zero modes of the edge excitations,
which describe the removal and adding of particles. 

 Additionally, we introduce a local back scattering impurity,
 which models the scattering of
 particles between the inner and outer edge.
  Since it has been shown that such a back scattering 
amplitude in a FQH bar is  strongly  renormalized and thus effectively
 temperature dependent\cite{1,moon}, this will be an additional source of the 
temperature dependence of the persistent  current.
  The back scattering amplitude of fractionally charged quasiparticles 
is found to be 
 enhanced as the temperature is lowered,
 so that  
one could even expect that the persistent current is depressed
as the temperature is lowered.
  This unusual expectation
 is  another motivation for  us to derive the
persistent current in a FQH- annulus with a weak impurity
 as a function of temperature in the following.

 In the next section we introduce the edge state theory 
 for the special geometry of a FQH- annulus.
 Then, we present the derivation of
 the persistent current and hall voltage in an isolated FQH- annulus,
 and discuss the result obtained.

 The more technical derivations are included in the appendices.
In appendix \ref{a0}, the hamiltonian of the gapless edge excitations 
is derived in a hydrodynamic model\cite{wen}. 
In appendix \ref{a1}, the quantization of the gapless edge excitations
and the bosonization of the quasiparticle operators is given.
In appendix \ref{a3} we present the path integral derivation of
the functional integral representation of the partition function.
In appendix \ref{a2}, the renormalization of the back scattering amplitude
is derived.
 In the last appendix \ref{a4} the 
 solution of series and integrals
 occuring in this article are listed.

 \section{ The FQH- annulus with a constriction}

The hamiltonian of the gapless edge excitations at the edge
of  the FQH- annulus  at filling factor $\nu=1/q$, q odd,  is given by
( see appendix \ref{a0} )
\begin{equation}\label{sign}
H_0=\frac{\pi }{\nu} \sum_{p=\pm} \int_0^{L_p} d x v_p (\rho_p(x)
 + p \nu \frac{\varphi}{ 2 \pi R_p})^2.
\end{equation}
 Here, $\rho_p(x)$ is the chiral electron density along the
edge, $L_p= 2 \pi R_p$ is the length of each edge.
 
 Note that the hamiltonian  depends on the flux $ \varphi$ piercing 
 the annulus additionally to the homogenous constant magnetic field B.
This accounts for the fact that  as the flux $\varphi$ is varied,
 the eigenstates are pushed up the outer edge and down the inner edge
 as first noted by Halperin\cite{halperin}, thus 
changing the total energy of the 
edge states.   Indeed, we see in Eq. (\ref{sign}),
 that the energy of the outer edge $p=+$, increases
when increasing the flux $\varphi = \phi/\phi_0$, while the energy of
 the inner edge, $p =-$  decreases.
 
 We  consider the weak back scattering case by an 
  additional term in the Hamiltonian due to the short range 
backscattering impurity  given by 
\begin{equation}
H_{BS} = \sum_{m=1} t_m \sum_{l=0}^{1/\nu -1} 
( \psi^{+ m}_+ (l L_+) \psi^{m}_- (l L_-) + c.c.).
\end{equation}
 Here, $m=1$ corresponds to backscattering 
of quasiparticles with charge $\nu$,
 while $m= 1/\nu$ describes the backscattering of
 chiral electrons. 
  The summation over l takes account of the fact 
 that the quasiparticle operator $\psi(x)$
is $1/\nu L_p$- periodic. Thus, the quasiparticle tunneling between 
the points $x=0,L_p,...,(1/\nu -1) L_p$ are distinguishable,
so that the  sum over them is explicit.          

 Next, the Hamiltonian can be quantized and the 
quasiparticle operators $\psi_p(x) $ can be bosonized in terms
of the edge magnetoplasmon modes.
 We  give the details in appendix
\ref{a1}.
With this microscopic theory of edge excitations, we
 are now able not only to obtain the periodicity 
 of thermodynamic properties, but also
 the magnitude of the persistent hall voltage and current
 in the annulus as a function of the 
flux
piercing it, its circumference and the
 temperature.

\section{The persistent hall voltage and current in a FQH- annulus}

 Having the full Hamiltonian and the complete algebra,
 we can derive the partition function of the FQH- annulus at finite 
temperature T.
By means of the path integral method
 the partition function can be expressed through a functional integral,
see appendix \ref{a3}. One obtains, 
\begin{equation}
Z = \int \prod_{\tau =0}^{\beta} 
\prod_{x=-L_p/2}^{L_p/2}  \prod_{p = \pm} d \phi_p(x)
\exp[ - S ]\mid_{N=const.}.
\end{equation}
  Here, the action S is given by ( see Eq. (\ref{action} )),
\begin{eqnarray}
S & = & \int_0^{\beta} d \tau \sum_{ p = \pm} \int_{-L_p/2}^{L_p/2} d x   
 ( p \frac{i}{4 \pi \nu} ( \partial_{\tau} \partial_x
\phi_p(x) ) \phi_p(x) \nonumber \\ &+& 
\frac{\pi v_p}{\nu} 
( p \frac{1}{2 \pi} \partial_x \phi_p(x) + p \nu \frac{\varphi}{ 2 \pi R_p} )^2 
) \nonumber \\
& + & \int_0^{\beta} d \tau \sum_{m=1} t_m 2 \sum_{l=0}^{1/\nu -1 } 
\cos m[  \phi_+(l L_+) - \phi_{-}(l L_-)
+  \frac{\pi \nu}{2}  N ] \nonumber \\
 & - & \int_0^{\beta} d \tau \sum_{ p = \pm}
p \frac{i}{4 \pi \nu} \phi_p(x) \partial_{\tau} \phi_p(-x) \mid_0^{L_p/2}.
\end{eqnarray}
 Note the last, new term  
which is due to the selfduality of the
 chiral edge field operators, and which is nonvanishing when the
 zero- modes are dynamic. See appendix \ref{a3} for details. 
We introduced   $N = N_+ + N_-$ as the total quasiparticle number on
the inner and outer edge of the FQH- annulus.
 The additional phase factor in the back scattering hamiltonian 
 is due to the anticommutation between chiral fermions
 on the inner and the outer edge.
 Since $ N_p = p \phi_p(L_p/2 ) - p \phi_p(- L_p/2)$, the 
 path integral derivation 
 could be performed as 
outlined in appendix \ref{a3}  even in the presence of these statistical
factors $C_p$ in the bosonized quasiparticle operators, Eqs. (\ref{qp},
 \ref{qp2} ).

  In order to perform the functional integrals, it is convenient 
to represent it in the momentum representation, Eq. (\ref{phip}),
 where we take the basis of magnetoplasmon modes of the annulus
 without impurity: 
\begin{equation}
 Z= \int \prod_{\tau = 0}^{\beta} d J (\tau) d \phi_J
\prod_{p = \pm}  \prod_{n \neq 0} d \phi_{p, n} (\tau ) 
\exp [ - S ]\mid_{N = const.}.
\end{equation}
 Here, we introduced the notation $J=N_+- N_-$
 with its conjugate, the phase difference between the two edges,
$\phi_J = \phi_{+ 0} - \phi_{- 0}$.  
The action in momentum space is: 
 \begin{eqnarray}\label{actionmomentum}
S&=& \int_0^{\beta} d \tau ( \sum_{p = \pm} ( \frac{i}{2}  \phi_J \partial_{\tau}
 J  
- \sum_{n>0} \frac{n}{\nu} (\partial_{\tau} \phi_{p,n} )
\phi_{+, -n} ) \nonumber \\
 &+& H [ \phi_{+}, \phi_{-} ] ),
\end{eqnarray}
where
\begin{equation} 
H[  \phi_p  ] = \sum_{p = \pm} H_p [\phi_p]
 + H_{BS}[\phi_{+}, \phi_{-}].
\end{equation}
The flux- dependent potential energy of edge p is diagonal in momentum,
\begin{equation}\label{fluxdependent}
H_p[ \phi_p] = \frac{ \nu v_p}{2 R_p} ( (N+pJ)/2 + p \varphi )^2 + \frac{1}{\nu} 
\frac{v_p}{R_p}\sum_{n > 0} n^2 \phi_{p, n} \phi_{p, -n},
\end{equation} 
while the back scattering term of the Hamiltonian mixes the 
 momentum Eigenstates of the clean FQH- annulus,
\begin{eqnarray}
H_{BS}[\phi_p] &=& \sum_{m=1}^{\infty} 2   t_m
\sum_{l=0}^{1/\nu-1}  \cos m [  \phi_J + 2 \pi \nu l N 
\nonumber \\
&+& \sum_{n \neq 0} ( \phi_{p,n} - \phi_{-p,n} ) + \frac{\pi}{2} \nu p   N ) ].
\end{eqnarray}
 Note that the action does not depend on $\phi_N = \phi_1 + \phi_2$,
 but, in the tunneling term,  on the phase difference   $\phi_J$,
 since the total particle number N is kept fixed, $\partial_{\tau} N=0$,
 in the isolated FQH- annulus.

 We make at this point an interesting observation:
 When there is a fractional total charge N, 
 the back scattering term of m quasiparticles
 is exactly zero due to the sum over l, 
 for all m besides 
 $m = 1/\nu n$ , n integer, corresponding to electron back scattering.
 
 Restricting N to multiples of $1/\nu$ in the following,
 which makes sense since the total charge of the
 FQH- annulus should not be fractional, we obtain thus:  
\begin{eqnarray}
H_{BS}[\phi_p] &=& \sum_{m=1}^{\infty} 2   t_m
1/\nu \cos m [  \phi_J 
\nonumber \\
&+& \sum_{n \neq 0} ( \phi_{p,n} - \phi_{-p,n} ) + \frac{\pi}{2} \nu p   N ) ].
\end{eqnarray}
  For simplicity, we substitute $ 2/\nu t_m$ by $t_m$ in the following.

 Now, we already can conclude from the action, Eq. (\ref{actionmomentum}),
  that the flux- periodicity of the partition function is $\phi_0$.
 To this end, we note that the only flux dependent term 
 of the action is the potential energy Eq. (\ref{fluxdependent} ).
  Thus, a  change of $\varphi$ by 1 does not change the partition 
  function 
 if there is a sum over families of different 
 numbers of fractionally charged quasiparticles $N_p$, 
 as noted by Thouless and Gefen\cite{5}, and 
which is explicit
  in the functional integral representation given  here.        

 Next, we  
study how
  the  back scattering amplitudes
$t_m$ are renormalized as a function of
 a typical energy, temperature or level spacing,
 $E = max(T, v_p/R_p)$,  by perturbative renormalization theory.
One finds as outlined in appendix \ref{a2}:
 \begin{equation}\label{reno}
 t_{m,eff}( E) = t_m (E/\Lambda)^{  m^2 \nu -1 } e^{ - \nu m^2 v/L/E/2 }.
 \end{equation}               
where $v/L = (v_+/L_+ + v_-/L_-)/2$.
 Thus, for temperatures  much exceeding the level spacing $v/R$,
the  quasiparticle backscattering amplitude, m=1,
 is enhanced when the temperature is
 lowered like a power law in 
 temperature, as found in Ref. \onlinecite{1,moon}, while higher order
backscattering amplitudes, $m^2 >1/\nu$ are irrelevant.
 However, we additionally 
  find an exponential prefactor, which becomes dominant,
 when the temperature approaches the finite level spacing $v/R$.
 This is reminiscent of the Coulomb blockade effect,
 where the charging energy is here the level spacing,
 the energy of an additional particle on the edge.  
  When the temperature is lowered below the level spacing,
 the renormalized tunneling amplitude 
 saturates at a temperature independent value,
 which  increases like a power law, when the level spacing is
 reduced. This is quite surprising, since intuitively,
one should think that the effect of a single back scattering
impurity  rather decreases, when the extension of the
 edge states is increased.

 Before proceeding in the calculation of the functional integrals,
 we consider first the expressions for the
persistent hall voltage and the persistent current.
 The thermodynamic persistent current is given by the
 flux derivative of the free energy:
\begin{equation}
I(\varphi) = T \frac{e}{2 \pi} \partial_{\varphi} \ln Z.
\end{equation}
 Using the above functional integral representation of the 
 partition function Z, we see that only the potential energy is flux 
dependent and we obtain:
\begin{eqnarray}
I( \varphi )  &=& - \nu \frac{e}{2\pi} \frac{v}{R} < T \int_0^{\beta} 
d \tau ( J(\tau)  + 2 \varphi) >
\nonumber \\
&-& \nu \frac{e}{4 \pi} ( \frac{v_+}{R_+} - \frac{v_-}{R_-} )
N.
\end{eqnarray} 
 where
\begin{equation}
< ... > = \int \prod_{\tau = 0}^{\beta}  d J d \phi_J \prod_p
 \prod_{n \neq 0} \phi_{p n} [ ... ] \exp [ - S ]/Z.
\end{equation}
 We see that there is even a finite persistent current without external 
 flux due to the difference in energy for
 a particle added to the inner or the outer edge. 

 Next, we define a  Hall voltage  
 by noting that 
 the voltage at a given imaginary time $\tau$
is related to the imaginary time derivative of
the phase difference $\phi_J$:
\begin{equation}
V(\tau)  = <1/(\nu e) \partial_{\tau} \phi_J > = - 1/( 2\nu e)
<\partial_{J} H(\tau)>. 
\end{equation}
 This is a quantity averaged over the circumference of the ring.
 We note that that  there are additionally local fluctuations
 of the edge to edge voltage. 
 Averaging over the imaginary time $\tau$, we obtain the 
 thermodynamic Hall voltage.
 Then, we find that this persistent Hall voltage is  
related  to the persistent current by a thermodynamic Hall conductance
\begin{equation}
\sigma_{xy} = \frac{I}{V} = \nu \frac{e^2}{2 \pi \hbar}.
\end{equation}

 It is noteworthy that this thermodynamic hall conductance
 is universal even in the presence of back scattering between the 
 edges. This is in contrast to the transport Hall conductance
which defers from this universal value as soon as there is back scattering,
 that is when there is no transversal localization between the edges.
  
 The reason for this drastically different behaviour 
 of the thermodynamic Hall conductance is clear:
 In the transport experiment a current is driven through the 
 sample externally and the quantum hall bar can only dynamically react
 to this external perturbation. Back scattering between the 
 edges reduces the Hall voltage but can not affect the 
 externally driven current, enhancing thus the Hall conductance. 
 In the isolated FQH- annulus on the other hand backscattering
  not only reduces the Hall voltage but also the thermodynamic current,
 so that the Hall conductance is unchanged.

 Now, we proceed in the explicit calculation of the persistent current
 in the presence of the single backscattering impurity.
We do an expansion in the scattering amplitudes $t_m$,
in order to be able to perform the functional integrals explicitly.

  The partition function has after expansion in
the tunneling amplitudes $t_m$ the form:
\begin{eqnarray}
Z &=& \sum_{w =0}^{\infty} \frac{1}{w!} \sum_{m_1...m_w=1}^{\infty} 
\prod_{l=1}^w t_{m_l} \int_0^{\beta} d \tau_1 ... d \tau_w 
\nonumber \\ &&
\int  \prod_{\tau = 0}^{\beta} d J(\tau)  
d \phi_J(\tau) \prod_p
\prod_{n \neq 0}  d \phi_{p, n}( \tau )  
\sum_{\alpha_1,..., \alpha_w = \pm}   
\nonumber \\ &&
\exp ( - S \mid_{t_m =0} ) 
\exp [ i \sum_{l=1}^w \alpha_l m_l ( \phi_J(\tau_l)
+ p \frac{\pi}{2} \nu  N 
\nonumber \\   &+&  
\sum_{n \neq 0 }  ( \phi_{+ n}( \tau_l) - 
\phi_{- n} ( \tau_l ) ) )]\mid_{N = const.}.
\end{eqnarray}   
 The integral over nonzero modes can be performed
by going to Matsubara frequency representation (see appendix \ref{a2} ), 
and gives a factor
 for the w-th order term  
which is independent on flux $ \varphi$:
\begin{equation}
\exp ( - \sum_{l,l' =1}^w \alpha_l \alpha_{l'} m_l m_{l'} D ( \tau_l - \tau_{l'} ) ),
\end{equation}
 where  one obtains after summation over the Matsubara frequencies,
\begin{equation}
D( \tau ) + D( - \tau) =   \nu \sum_p \sum_{n>0} \frac{1}{n} 
\frac{\cosh ( n v_p/(2 R_p T) ( 1 - T
\mid \tau \mid ))}{\sinh ( n v_p/(2 R_p T) )}. 
\end{equation}
 Since the edge magnetoplasmons were considered in harmonic
approximation, it is not surprising that this 
 factor due to the gapless  nonzero edge modes
resembles the Debye- Waller- factor, describing the 
 thermal damping of periodic structures.

 Next we can do the remaining integrals over 0-modes:
\begin{eqnarray}
Z_w^0 & = & \int \prod_{\tau =0}^{\beta} d J(\tau ) d \phi_J(\tau )
\nonumber \\
& &\exp [- \int_0^{\beta} d \tau ( \frac{i}{2} \phi_J \partial_{\tau}
J 
 +  \frac{ \nu v }{ R} (\frac{1}{2} J +   \varphi )^2 
+ \frac{\nu v}{4 R} N^2 
\nonumber \\ 
&-& i \sum_{l=1}^w \alpha_l m_l 
( \phi_J ( \tau_l ) + \pi/2 \nu p  N ))].
\end{eqnarray}

Performing first the integral over the phase difference   
 $\phi_J$ gives a delta- function, which ensures
 that the change of the particle number difference $ J $ on one edge  happens
 only at the tunneling events $\tau_l$:
$\partial_{\tau} J =  2 \sum_{l=1}^w \alpha_l m_l \delta ( \tau_l - \tau) $.
 Thus, the difference in
particle number  betweeen the two edges  is constrained to be    
\begin{equation}
J(\tau) = \bar{J} + 
2  \sum_{l=1}^w \alpha_l m_l \theta_{step} ( \tau - \tau_l),
\end{equation}
where $\bar{J}$ is an even integer. 
Note that the periodic boundary conditions $ J(\tau + \beta ) = J(\tau) $
 require that 
\begin{equation}
\label{neutral}
 \sum_{l=1}^w \alpha_l m_l= 0,
\end{equation}
 exactly, 
 which 
 means that the total change in charge on either edge, after w tunneling
 events,  
has to be  zero. This is a consequence
 of the fact that the partition function
 is a trace, that is, one must come back to the same state when going in the
 path integral from $\tau =0$ to $\tau = \beta$.    
 Note that while this condition makes the first order term  
 vanish this is not true for odd terms in general.
 For example  in the third order term, $w=3$,
 the condition Eq. (\ref{neutral}) can be fullfilled
 with f.e. $ m_1 = m_2 =1, m_3 =2, \alpha_1 = \alpha_2 = -\alpha_3 $,
 which corresponds to twice a $2k_F $ scatttering from edge $p=+$
 to $p=-$ followed by a $4 k_F$ scattering back.

  Now, also the integral over $J$ is performed,
 up to a summation over the integer numbers $\bar{J}$.
 This is a summation and not a continous integral, since only multiples 
of the charge of a quasiparticle can be added or removed from one edge,
 due to the edge to edge back scattering which does not allow
 a continous transfer of charge.   
 In fact, the persistent current would vanish exactly, if 
there would be a continous integral rather than
 a discrete sum over $\bar{J}$.
 Thus, for the existence of a persistent current it seems   essential
 that only a discrete amount of charge can tunnel between the edges.

 Note that at this point the factor 
\begin{equation}
\exp( - i \frac{\pi}{2} \nu p \sum_{l=1}^w \alpha_l
 m_l  N ),
\end{equation}
originating from the
 anticommutation between the chiral fermions on the inner and the outer edge
 dropped out because of the condition Eq. (\ref{neutral}),
 so that we can conclude that 
 it is not essential for the partition function,
 if the  anticommutation
 between the edges is taken into account or not.

 We obtain for the partition function 
 with the condition of neutrality Eq. ( \ref{neutral} )
 and transforming to $x_l = T \tau_l$:
\begin{eqnarray}
Z &=& \sum_{\bar{J} even} \exp ( - \frac{\nu v}{4 T R} ( \bar{J} + 
2 \varphi )^2) 
\nonumber \\ &&
\sum_{w=0}^{\infty} \sum_{m_1,...,m_w} \sum_{\alpha_1,...,\alpha_w = \pm}
\mid_{Eq. \ref{neutral}} 
\prod_{l=1}^w ( \frac{t_{m_l}}{T} \exp( - m_l^2 D(0) )
) 
\nonumber \\ &&
\int_0^1 d x_1 \int_0^{x_1} d x_2 ... d x_w
\exp ( - \sum_{l \neq l'} \alpha_l \alpha_{l'} m_l m_{l'} D ( x_l -x_{l'} ))
\nonumber \\&&
\exp ( \sum_{l=1}^w \alpha_l m_l x_l \frac{\nu v}{T R} ( \bar{J} + 2 \varphi
- \sum_{l' < l} \alpha_{l'} m_{l'} ))
\end{eqnarray}

 While 
 similar integrals could be performed exactly previously, ( see Ref. 
\onlinecite{exact} ), this is complicated here 

1. by the 
exponential factor originating from the dynamics of the 0-modes 
which is not present in the macroscopic limit $R \rightarrow \infty$.

2. the fact that there are relevant mixed terms
   like f.e. $ w=3, m_1=m_2=1, m_3=2, \alpha_1=\alpha_2=-\alpha_3$.

 Therefore, we had to give up looking for an exact solution 
 at this point and restrict us now to the second order 
 correction $ w=2$.  
 In fact, as we found  above by perturbative renormalization
 theory, Eq. (\ref{reno}), in a finite system
 the quasiparticle back scattering 
 amplitude saturates  quickly as the temperature is lowered,
 even for temperatures larger then the level spacing 
$ T > v/R$ due to a blockade effect.
 Thus, we may hope to catch the essential physics 
 of the weak backscattering limit $t_m/\Lambda \ll 1$ 
by restricting
 us now to 2 nd order perturbation theory.

 The persistent current  in the annulus
 to second order in the back
 scattering amplitudes $t_m$ is obtained to be given by, 
 \begin{eqnarray}\label{pc}
I & = & - \nu \frac{e}{2 \pi} \frac{v}{R} ( <  \bar{J}  >_{\bar{J}} + 2
\varphi 
 +   \sum_{m=1}^{\infty} \tilde{t}_m^2 
\nonumber \\
&&( < \bar{J} 
P_m(\bar{J}+ 2 \varphi) >_{\bar{J}}
- < \bar{J} >_{\bar{J}}
 < P_m ( \bar{J} + 2 \varphi) >_{ \bar{J}} ))
\nonumber \\ &&
+ \frac{e}{2 \pi} T \sum_{m=1}^{\infty} 
\tilde{t}_m^2   <\partial_{\varphi} P(\bar{J} + 2 \varphi)>_{\bar{J}}
\nonumber \\
&& - \nu \frac{e}{4 \pi} (\frac{v_+}{R_+} - \frac{v_-}{R_-} ) N,  
\end{eqnarray}
 where we introduced the function
\begin{eqnarray}
P_m(z) &=& \int_0^{1} d x e^{m^2 (D(x) + D(-x))} 
\nonumber \\ &&
2 \cosh ( \frac{ \nu v}{ T R} m z x )
 \exp( - \frac{ \nu v}{T  R} m^2 x ).    
\end{eqnarray}
and the renormalized tunneling amplitude
\begin{equation}
\tilde{t}_m = \frac{t_m}{T} e^{- m^2 D(0)}.
\end{equation}
 The average is taken with respect to the unperturbed system:
\begin{equation}
<...>_{\bar{J}}
 = \sum_{\bar{J} even} \exp( - \frac{\nu v}{4 T R} ( \bar{J} + 
2 \varphi )^2)[...]/Z^{(0)}. 
\end{equation}
where $Z^{(0)}$ is the partition function of the unperturbed system.

 We see that 
 the persistent current is exponentially small when the 
temperature T exceeds the level spacing $v/R$. Therefore,
 the power law increase with temperature of the back scattering amplitude 
 in this temperature regime has no effect on the 
magnitude of the persistent current.
  We see this more clearly by considering the first harmonic
 of the current, which is for $T > v/R$ given by, keeping only
 relevant terms,
\begin{equation}
I_1(T) = - 2 e T \exp ( - \frac{\pi^2}{\nu} \frac{T}{v/R}) 
( 1 - t_1^2 (\frac{\pi}{\Lambda})^{2 \nu} T^{2 (\nu-1)} B(T)). 
\end{equation}
 where
\begin{equation}
B(T) = \int_0^1 d x \frac{ (\sin  \pi x)^2 }{(2 \sin \pi/2 x )^{2 \nu}} 
\exp ( \frac{\nu v}{R T} x^2 ).
\end{equation}
  Thus, the exponential prefactor is dominating down to temperatures
T equal to the level spacing, and a decrease in the persistent current
 when lowering the temperature can not be observed.
 
 When the temperature is  below the level spacing, 
 the magnitude of the persistent current increases
 and higher harmnics become important,
 so that the shape as a function of flux changes 
 from a sinusoidal to a sawtooth function.
 
 In that limit, 
\begin{equation}
\tilde{t}_m^2 P_m(z) = \frac{v/R}{T} t_m^2  (2\Lambda )^{- 2 m^2 \nu} 
( \frac{v}{R} )^{2(\nu m^2 -1)} B_m(z, \frac{v/R}{T}),
\end{equation}
where
\begin{equation}
B_m(z, \frac{v/R}{T}) =
\int_0^{v/R/T} d y
\frac{ 2 \cosh ( \nu m z y) }{(2 \sinh (y/4))^{2 \nu}} \exp ( -
  \frac{\nu}{2} y ).
\end{equation}
 Now, we consider the flux interval $-1/2 < \varphi < 1/2$, and the result 
 is understood to be extended periodically. 
 For $ T \ll v/R$ and  $ \mid \varphi \mid \ll 1/2$,
 the current is well approximated by
using $< \bar{J}> =0$. Then, keeping relevant terms, only, one obtains,
\begin{eqnarray}
I( \varphi )  &=& - \nu \frac{e}{\pi} \frac{v}{R} \varphi 
 \nonumber \\
& + & \nu \frac{e}{2 \pi} (\frac{v}{R})^{2 \nu -1} 
t_1^2 ( 2 \Lambda)^{-2 \nu}   
 \partial_{\varphi} B_1(\varphi, \infty).
\end{eqnarray}
 Therefore, we can conclude that the
  persistent current is temperature
 independent for temperatures far below the level spacing, and not too close
to the turning points $\mid \varphi \mid =1/2$.
 Furthermore, 
 the amplitude of the back scattering correction increases like a power
 law with decreasing level spacing, $\sim (v/R)^{2 \nu -1}$,
 thus reducing the persistent current. 

 In Fig. 2 we plot the persistent current, Eq. ( \ref{pc} )
 for several temperatures below the level spacing 
and compare it with the 
 one obtained for the  FQH- annulus without constriction,
 plotted in Fig. 3.
 We observe that only at very low temperatures a hundred
 times smaller than the level spacing the reduction due to back scattering
 becomes appreciable. The peculiar behavior of
 the current for $T/(v/R)= 1/100$ and flux close to $\pm 1/2$,
 indicates a breakdown of second order perturbation theory in this 
 regime.  

In summary,
 the persistent current in a 
FQH- annulus  with a
 weak backscattering
 impurity was calculated. 
    It was  proven  
 that the persistent current is flux periodic with period $ \phi_0$
 due to the possibility  of fractionally charged quasiparticles
 to scatter between the edges of the FQH- annulus.
  At temperatures T exceeding  the level spacing $v/R$,
 the current is exponentially small, and the back scattering correction,
 which increase like a power law, when reducing the temperature,
 has no effect on the magnitude of the current.
 
  At temperatures far below the level spacing, the current becomes 
 temperature independent, and the back scattering correction 
 increases like a power law, when the level spacing is
 reduced.
 
   The persistent current is found to be proportional
 to the fractional charge $\nu e$ of the quasiparticles, see Eq. (\ref{pc}).

 We also found that one can define a thermodynamic Hall conductance  
  which has the universal value $ \sigma_{xy} = \nu e^2/( 2 \pi \hbar )$,
 even in the presence of back scattering between the edges.

  Experimentally, the total magnetization of a FQH- annulus
 could be measured, where a combination of low temperatures
 and small samples is preferable in order to observe
 a signal due to the persistent current. 
  The magnitude of the Hall voltage calculated above is presumably only 
 accesible experimentally, if it is combined with a simultanous
 measurement of the magnetization, thus making it possible to
 subtract externally produced voltage fluctuations  
   which cause Hall current fluctuations in the ring.

 An interesting question is, if the persistent current 
 changes its magnitude, when the filling factor is
 changed away from $\nu =1/q$, or remains on a plateau,
 reminiscent of the quantum Hall effect\cite{avishai,kim}.
  Since the method employed here relies only on the incompressibilty
 of the bulk of the FQH- annulus we expect the result to hold 
 as long as there is a gap for bulk excitations.
 At temperatures below the level spacing $v/R$,
 there is a dependence on magnetic field due to 
 the relation to the edge velocity: $v = E/B$, where E is
 the gradient of the edge potential. 
 Additionally there might be a strong hysteresis effect, since
 when changing the total magnetic field it might be impossible
 to fix the additional flux $\varphi$ through the ring,
 making it hard to find the functional dependence on
 the magnetic field,  Eq. $\ref{pc}$.

\section{Acknowledgement}
The author has benefited
during a stay at the Condensed Matter Physics Department 
of the Weizmann Institute
 from numerous discussions.
 Particular thanks to Yuval Gefen, Dror Orgad, Yuval Oreg, Alex Kamenev, 
and Alexander M. Finkel'stein.
The support from the Minerva Foundation 
and the Max- Planck- Institute for Physics of Complex Systems
as well as  from 
a fund of the Israeli Academy of Science, the German Israeli Foundation
(GIF),
and the U.S.- Israel Science Foundation (BSF) is gratefully acknowledged.
\appendix

\section{The Hamiltonian of the gapless edge excitations}
\label{a0} 
The effective Hamiltonian of the gapless edge excitations
 can be found as the gapless excitations of the two- dimensional
 Chern- Simons- theory describing the interacting electrons
 in the strong magnetic field in the 2- dimensional annulus\cite{wen,dror}. 
 Instead, we give here a hydrodynamic  derivation of the Hamiltonian
 of gapless edge excitations, using only the incompressibilty of the
 bulk system and the existence of a sharp edge\cite{wen}.  
 
This derivation is valid provided that

 i) the edge is sharp, that is  
     the thickness over which the potential
     varies by the amount of the cyclotron frequency
  $\omega_c= eB/m$ is smaller than
    the magnetic length.  

 ii) the long range interaction is small enough 
      ( the gapless edge excitations have a dispersion
      $E_k = v(k) k$ where $v(k)$ is constant 
      for $ k > 1/l_B \exp ( - \pi/ \nu \omega_c/ ( e^2/\epsilon/l_B))$.
      where $l_B$ is the magnetic length, 
      given by $l_B^2 = c/e/B $, the dielectrical constant 
is typically of the order of 
$\epsilon \approx 10$,
 and $\omega_c$ the 
       cyclotron frequency\cite{dror}.
Additionally, interedge interactions are disregarded in the
 following\cite{oreg}.
            
  Now, we outline the hydrodynamic derivation.
Consider a  quantum hall bar
having an edge potential which is for small deviations $y_p$ from 
the edges  linear: $E_p y_p$. 
 Then we can  derive an effectively one- dimensional Hamiltonian by
measuring  the energy of  deviations 
in the shape of the incompressible bulk 
at the  edge.    
We will measure those deviations normal to the surface of the quantum hall
bar.
Thus, $y_-$ is positive if the FQH- liquid rises beyond inner edge, while 
$y_+$ is positive if the FQH- liquid moves up the  outer edge.
This way, we find for the energy of the quantum hall bar due to deviations 
from the edge of minimal energy: 

\begin{equation}
H_0 = \sum_{p = \pm} \int_{  edge p  }  d s \int_0^{h_p(s)} d y_p  e n E_p
 y_p.
\end{equation}

 Note, that the magnetic field perpendicular to the annulus causes a drift 
velocity  along the edge given by $v_p=\nu e E_p/ 2 \pi n$ where $\nu=\phi_0 n/B$
 is the filling factor, the amount of particles
 per flux quantum $\phi_0=hc/e$,
with n being the particle density. Introducing $\rho_p(s)=n h_p(s)$,
where $p = \pm$,  which have the dimensions of  1-dimensional
densities and can be viewed as the  particle densities along the edges,
 we obtain in harmonic approximation:
\begin{equation}
H_0=\frac{\pi }{\nu} \sum_{p=\pm} v_p \int_{edge p} d s \rho_p(s)^2.
 \label{ham0}.
\end{equation}

\section{Quantization}
\label{a1}

In this appendix we quantize  the Hamiltonian of the gapless
 edge excitations, and obtain the algebra of the
edge excitations.
The system is characterized by 
a continuity equation for the density of chiral 
electrons on each edge of length $L_p$. 
\begin{equation}\label{cont}
\left(\partial_t + p 
v_p \partial_{x} \right) \rho_p(x,t)
= U_p(x,t),
\end{equation}
where $p=\pm$ enumerates the edges and
the source and drain function $U_p(t)$ is due to the possibility  
of back scattering from edge to edge across the bulk
at an  impurity.

The quasiparticle operators $\psi_{ p}(x)$, removing a quasiparticle of charge
$\nu e$ from edge $p=\pm$,
 can be related to the
chiral electron  density at the edge, $\rho_p(x)$ by the bosonization technique.
\begin{equation}\label{qp}
\psi_p(x)=C_p \exp (i \phi_p(x) ),  
\end{equation}
where
\begin{equation}\label{dens}
\rho_p(x)=p\frac{1}{2 \pi } \partial_{x} \phi_p(x),
\end{equation}
with $p=\pm$ denoting the edges of opposite chirality. 
$C_p$ is an operator ensuring anticommutation between chiral electron operators
 on opposite edges, and will be specified below.
Since $\rho_p(x)$ is the electron density along the edge p,
$\phi_+(x)= 2 \pi \int^{x} d x' \rho_+(x')$ 
counts $2 \pi$ times 
the number  of chiral electrons at positions on the right going
edge, +,  smaller than $x$,
while
$\phi_-(x) = 2 \pi \int_x d x' \rho_-(x')$
 counts accordingly $ 2 \pi$ times the number of 
particles on positions $x' > x$ on 
the left going edge. 
 Here, the difference in the definition for the two edges 
 arises from the fact that we have chosen the same coordinate
 system for both edges, whose chirality is opposite, however. 

Besides the linear dependence on $x$ due to a finite number
of quasiparticles on an edge, $N_p$, we can use a Fourier expansion
 for $\phi_p(x)$
and get:
\begin{equation}\label{phip}
\phi_p(x)= \phi_{p 0} + p \nu N_p  x/R_p  + 
\sum_{n \neq 0} \phi_{p n} e^{i n  x/R_p}.
\end{equation}

The density of chiral electrons 
is taken to be periodic: $\rho_p(x + L_p)=\rho_p(x)$,
 and its Fourier expansion is
\begin{equation}\label{dens2}
\rho_p(x)= \nu \frac{N_p}{ 2 \pi R_p}
 + \sum_{n \neq 0}
\rho_p(n) e^{i n  x/R_p}.
\end{equation}

 The Fourier components of the chiral electron 
density 
$\rho_p(n)$ are related to the Fourier components of the
phase operator 
$\phi_p(n)$  
as 
\begin{equation} \label{densn}
\rho_p(n)= i p \frac{n}{2 \pi R_p}  \phi_p(n)
\end{equation}
 for $ n \neq 0$.

{\it Quantization of the nonzero modes}
 
 Now, from
 the classical Hamiltonian equations 
combined with
 the continuity equation Eq. (\ref{cont}),
we can obtain after quantization the 
 algebra for the density operators
in momentum space.
  
 We first consider the Hamiltonian equations which define
for the Fourier components of the field, $\phi_{p,n}$, their conjugate fields
$P_{p,n}$:   
\begin{equation}
\partial_t P_{p,n} = -\frac{\partial H}{\partial \phi_{p,n}}.
\end{equation} 
The Hamiltonian Eq. (\ref{sign} )
of the edge excitations is in momentum representation,
 using Eqs. ( \ref{densn}, \ref{dens2} ) , found to be given by 
\begin{equation}
H_0= \sum_{p=\pm}
\frac{ \nu v_p}{   2 R_p}  (N_p + p \varphi)^2
 +  \frac{1}{ \nu } \sum_{p=\pm} \frac{v_p}{R_p}
\sum_{ n > 0} n^2 \phi_{p,n} \phi_{p,-n}.
\end{equation}
Thus, we find that the Hamiltonian equation becomes
\begin{equation}
\partial_t P_{p,n} = - \frac{2 \pi v_p}{\nu } n^2 \phi_{p,-n}(t)
-\frac{\partial H_{BS} }{ \partial \phi_{p,n}}.
\end{equation}
 Here, $H_{BS}$ is the part of the Hamiltonian which describes backscattering
of quasiparticles from an impurity.   
Combining the continuity equation Eq. (\ref{cont}) 
with this Hamiltonian equation, we thus can 
identify the conjugate field as  
\begin{equation}
P_{p,n} = i p \frac{ n}{\nu} \phi_{p,-n}(t), 
\end{equation}
provided that the source and drain term in the continuity equation
is related to the  impurity Hamiltonian as
\begin{equation}
U_{p,-n}(t) = - \frac{\nu}{2 \pi R_p} \frac{\partial}{\partial \phi_{p,n} } H_{BS}.
\end{equation}
 Having obtained a complete set of conjugate fields, $\phi_{p,n}, P_{p,n}$,
for $ n > 0$,  
 we can quantize the thus identified conjugate fields
and get
\begin{equation} 
[p_{p,n},\phi_{p',n'}] = i \delta_{n n'} \delta_{p , p'},
\end{equation}
 from which follows for the Fourier components of the density operators
\begin{equation} \label{com1}
 \left[ \rho_{p,n},\rho_{p',n'} \right] =
  - p \frac{n \nu}{L_p^2} \delta_{n,-n'}\delta_{p, p'},
\end{equation}
 and for the Fourier components of the phase operators,
\begin{equation}
\left[\phi_{p,n},\phi_{p',n'}\right] =  - p  \frac{\nu}{ n} \delta_{n,-n'}
\delta_{p,p'}.
\end{equation}
 Note that  the Fourier components of field operators  
from opposite edges $p = \pm$ do commute.  This is a consequence 
of our choice of the basis as the set of Eigen modes on each separate
edge.

{\it quantization of the 0- modes:}

Special care has to be taken of the zero modes, $n=0$.
While the nonzero modes $\rho_{p,n}$ describe  fluctuations 
of the edge which do not change the total charge on each edge,
and which have a linear dispersion,
 $N_p$  is the total number operator of quasiparticles 
with charge $q=\nu e$ on edge $ p = \pm$
and its conjugate $\phi_{p 0}$ is related to the 
ladder operator, adding a particle to the edge p, as
$\exp ( i/ \nu  \phi_{p 0})$
 We do obtain the algebra of the quantized 0- modes 
by demanding that 

1. the ladder operator indeed adds a quasiparticle  to edge p:
\begin{equation}
\exp(i \phi_{p 0}) N_p = (N_p+1) \exp(i \phi_{p 0}).
\end{equation}

2. The
 bosonized Fermion operators obey Fermi statistics, and thus anticommute:
 \begin{equation}
\{\psi_p(x),\psi_{p'}(x')\} = 0.
\end{equation}

{\it Anticommutation between Fermi operators  on the same  edge:}

 With $\psi_p^{1/\nu}(x)
 = C_p^{1/\nu} \exp( i (1/\nu ) \phi_p(x) )$  follows,
with  the Baker- Haussdorff formula for equal edges, $p = p'$:
\begin{eqnarray}&&
(\exp(- \frac{1}{2 \nu^2} [ \phi_p(x), \phi_{p}(x') ] )
 + \exp(  \frac{1}{2 \nu^2} [ \phi_p(x), \phi_{p}(x') ]) )
\nonumber \\ 
&& C_p^{2/\nu} \exp( i \frac{1}{\nu} (\phi_p(x) + \phi_{p}(x') ) = 0,
\end{eqnarray}
which means that
\begin{equation} \label{com2}
[ \phi_p(x), \phi_{p}(x') ] = - i  p  \nu^2 (2 s +1 ) \pi 
sgn ( x - x'). 
\end{equation}
 Thus, the condition of anticommutation of the
 bosonized fermion operators still leaves the freedom to choose the integer $s$.
  This freedom, however is lost when we demand that 

1 . $\psi_p^{1/\nu}(x)$ changes the
 density $\rho_p(x')$ by $-1/L \delta( x-x') $, corresponding to a removal
 of a charge  of an electron
and

 2. the phase accumulated under the exchange of two quasiparticles 
at the edge
is $ \pi \nu $ as in the bulk of the FQH liquid.
 Thus, we obtain the condition $ 2 s + 1= 1/\nu
 $.

 Note that the phase accumulated under exchange of two edge fermions is then
 $1/\nu \pi$ rather than $\pi$ as for usual fermions.

  Now, we will use  these commutation relations 
in real space 
to find the commutation relations of the zero modes,
which is possible, since we do know already the commutation relations
of the nonzero modes which we had derived
above from the Hamiltonian equations combined with
the continuity equation.

For $p=p'$, it follows from Eqs. (\ref{phip},\ref{com1},\ref{com2} ) that 
\begin{equation} \label{zero1}
[N_p,\phi_{p 0}]=i.
\end{equation}

{\it anticommutation between fermi operators defined on opposite edges:}

Next, demanding the anticommutation of fermion operators 
defined on opposite edges $p \neq p'$,
it follows 
\begin{equation}\label{qp2}
C_p = \exp ( i \nu^2 ( 2s +1) p \frac{\pi}{2} N_{-p} ).
\end{equation} 
 Demanding additionally that the exchange of two quasiparticles on
different edges gives a phase $\pm \nu \pi$,
 the freedom in s is lifted: $2s +1 = 1/\nu$. 

{\it Algebra of $\phi$ and $\theta$: } 

Having now the complete   set of commutation relations in real
and in momentum space we now look for conjugate 
operators which are not self dual in real space, which means that 
we look for operators which do commute taken at different points in space.

One finds, using Eq. (\ref{com2}),
that these nonself dual operators are the symmetric and antisymmetric 
combinations of the operators on opposite edges:
\begin{equation}\label{phi}
\phi (x) = \phi_1 (x) + \phi_2(x),
\end{equation}
and
\begin{equation}\label{theta}
\theta (x) = \phi_1(x) - \phi_2(x).
\end{equation}

We find that $\phi$ and $\theta$ do not commute with each other:
\begin{equation}
[\phi(x),\theta(x')]
= - 4 \pi \nu i \theta_{step} (x - x'). 
\end{equation}
However, each is not self dual: 
\begin{equation}
[\phi(x), \phi(x ') ] = 0,
\end{equation}
and
\begin{equation}
[\theta(x),\theta(x')] = 0.
\end{equation}

\section{Chiral Action}
\label{a3}

 Here we give the path integral derivation of the functional integral representation
 of the partition function of a chiral edge state of an incompressible FQH liquid. 
  This turns out to be nontrivial due to the chirality of the edge states,
 the selfduality of the chiral edge field operators, and 
the dynamics of the 0- modes in a finite FQH- annulus.
 We sketch this derivation for the case of the FQH- annulus  in
 the following.

 The partition function is given by
\begin{equation}
Z = Tr [ e^{ - \beta \hat{H} } ]
\end{equation}
Dividing the inverse temperature $ \beta$ into 
infinitesimally small imaginary time slices
 $\epsilon$ with $ M \epsilon = \beta$  one obtains
\begin{eqnarray}
Z &=& Tr \prod_{j=0}^M e^{- \epsilon \hat{H} } \nonumber \\
 &=& Tr \prod_{j=0}^M : e^{ - \epsilon \hat{H} } :. 
\end{eqnarray}
  Here $ : ... :$ denotes normal ordering, and $\hat{H} $ is
 assumed to be written in terms of conjugate operators
 $\hat{\phi}, \hat{\phi}^*$.  The correction to the normal ordered
 factors are of order $ \epsilon^2$ each and can therefore 
 be neglected in the limit $ \epsilon \rightarrow 0$,
$ M \rightarrow \infty$, while $ M \epsilon = \beta$ is 
 kept fixed to the inverse temperature $\beta$.\cite{negele}
  
 For the chiral edge, there
 arises the problem in the derivation of the functional
integral, that the operators defined on one chiral edge
are selfdual, see appendix \ref{a1},
so that one can not straightforwardly apply the
 Feynman path integral method. 
 Therefore, one has to find nonselfdual combinations of these
operators and their conjugates. One possible choice
are the combinations $\phi$ and $\theta$ of the
fields defined on opposite edges, Eqs. (\ref{phi},\ref{theta}).
However, we find it more convenient to
divide each chiral edge into two parts,  and form nonselfdual
operators as combinations of operators defined in these two parts.
We first sketch the derivation for a single chiral edge
of length $L_p$,
and can then easily extend the result to several edges. 

The combinations
 $\hat{\phi}_{s,a} = \hat{\phi}_p(x) \pm \hat{\phi}_p( -x ) $
of the chiral fields $\hat{\phi}_p(x)$ 
are not selfdual for $ 0 \leq  x \leq L_p/2$:
\begin{equation}
[ \hat{\phi}_{s,a} (x), \hat{\phi}_{s,a} (x') ] = 0.
\end{equation}     
 while they do not commute with each other,
\begin{equation}
[ \hat{\phi}_s(x), \hat{\phi}_a(x') ] = p
4 \pi i \nu \theta_{step} (x'-x).
\end{equation}
 
 Now, we can continue 
in the derivation of the functional integral in terms of the
Eigen values of the conjugate opertaors 
\begin{equation}
\hat{\phi}(x) = \hat{\phi}_s(x),
\end{equation}
\begin{equation}
\hat{\phi}^*(x) = \frac{1}{4 \pi i \nu^2 (2 s +1 )}
 \partial_x\hat{\phi}_a(x).
\end{equation}
with
\begin{equation}
[\hat{\phi}(x), \hat{\phi}^*(x')] = \delta(x -x')
\end{equation}
The respective coherent states are
\begin{equation}
\mid \phi > = e^{\int_0^{L_p/2} d x \phi(x) \hat{\phi}^*(x) } \mid 0 >,
\end{equation}
 which satisfy the completeness relation,
\begin{equation}
\int \prod_{x=0}^{L_p} \frac{ d \phi^*(x) d \phi(x) }{ 2 \pi i} 
e^{ - \int_0^{L_p/2} d x \phi^*(x) \phi (x) }
\mid \phi > < \phi \mid = 1.
\end{equation}  

 Then, the partition function is obtained
 as
\begin{eqnarray}
Z &=& \int \prod_{j=0}^M  \prod_{x=0}^{L_p/2} \frac{ d \phi^{(j) *}(x)
 d \phi^{(j)} (x) }{ 2 \pi i}
e^{ - \sum_{j=0}^M \int_0^{L_p/2} d x \phi^{(j) *}(x) \phi^{(j)}(x) }
\nonumber \\ &&
\prod_{j=0}^M < \phi^{(j+1)} \mid : e^{ -\epsilon \hat{H} [ \hat{\phi},
\hat{\phi}^* ] } : \mid \phi^{(j)} >.
\end{eqnarray}
 Because of the normal ordering, the
 matrix elements in coherent state representation
 at each time slice are easily evaluated as
\begin{eqnarray}
&&< \phi^{(j+1)} \mid : e^{ - \epsilon \hat{ H} [ \hat{ \phi } , 
\hat{ \phi}^*  ]} : \mid \phi^{(j)} >  
\nonumber \\
 &=& < \phi^{(j+1)} \mid \phi^{(j)} >  
e^{- \epsilon H [ \phi^{(j)}, \phi^{(j+1) *} ]}.
\end{eqnarray}
 
  Having thus the functional integral representation one can do a 
 transformation back to the fields defined on a single point,
$\phi_p(x)$, 
yielding the correct effective action of one chiral edge,
which can be checked by variation of this effective action,
giving the correct equations of motion,

\begin{equation} \label{actionp}
S = S_0 - \int_0^{\beta} d \tau p \frac{i}{4 \pi \nu } \phi_p(x)
\partial_{\tau} \phi_p(- x) \mid_0^{L_p/2}, 
\end{equation}
 where $\phi_p(x)$ is the chiral field defined on one edge.
Here, $S_0$ is the action one would obtain, disregarding  
the selfdual nature of the operators on a chiral edge:
\begin{equation}
S_0 = \int_0^{\beta} d \tau
( \int_{-L_p/2}^{L_p/2} d x  p \frac{i}{4 \pi  \nu } 
(\partial_{\tau} \partial_{x} \phi_p(x)  ) \phi_p(-x)
+ H[ \phi_p(x) ] ).
\end{equation}
   Disregarding  the dynamics of the 0 - modes, one would
indeed obtain $S =S_0$. However, including the dynamics
of the 0- modes, which are 
the particle number $N_p$ and its phase conjugate
$\phi_{p 0}$, 
 $S_0$ itself contains a
term which couples the 0- modes to the nonzero- modes,
giving an unphysical term to the equations of motions,
as obtained by variation of the action $S_0$.
 The additional term appearing in Eq. (\ref{actionp})
cancels exactly this unphysical term contained in $S_0$,
thus yielding the correct equations of motion.
  
 Having the partition function of a single chiral edge, we can 
easily give the one for a FQH- annulus with two edges:
Noting that the Hamiltonian can now be written as
a functional of $\hat{\phi}_p$, $p = \pm$, and that $\hat{\phi}_+(x)$
and $\hat{\phi}_-(x)$ do commute with each other,
we obtain for the total action of the FQH- annulus
\begin{equation} \label{action}
S = S_0 - \int_0^{\beta} d \tau \sum_{p= \pm} p \frac{i}{4 \pi \nu } 
 \phi_p(x) \partial_{\tau} \phi_p(-x)\mid_0^{L_p/2},
\end{equation}
where
\begin{eqnarray} 
S_0 &=& \int_0^{\beta} d \tau ( \sum_{p=\pm} p \frac{i}{4 \pi \nu}
\int_{-L_p/2}^{L_p/2}
 d x \partial_{\tau} \partial_x \phi_p(x) ) \phi_p(x) 
\nonumber \\&
+& H[ \phi_+(x),
\phi_-(x) ] )
\end{eqnarray}
 
 \section{Renormalization}
 \label{a2} 

 Here we give a derivation 
 in perturbation theory
of the Renormalization flow equations for the
 backscattering amplitude.

   Unlike  the route of derivation in real space taken by Kane and Fisher
\cite{1}
 we find it more convenient to give the derivation in momentum space.  

 Since the scattering term is local in real space, the matrix in
 the Gaussian integral is nondiagonal in momentum space.
 However, we show in the following that to first order in the
tunneling amplitudes $t_m$, only the diagonal elements of this
 matrix contribute to the renormalization.
  Consider a Gaussian integral of the form
\begin{equation}
\int \prod_i d z_i dz^*_i \exp ( -\sum_{ij} z_i A_{ij} z^*_j ) 
= 1/ det A = \exp ( - Tr \ln  A )
\end{equation}
When $A$ has the form $A = A_0 + t B$ where $t$ is a small parameter,
one can expand in t and find to first order in t:
\begin{equation}
\exp ( - Tr \ln A_0 - Tr t \frac{B}{A_0} )
\end{equation}
 When $A_0$ is a diagonal matrix, we find that, indeed $t$ is
 renormalized by the diagonal matrix elements of $B$, only,
 which we wanted to prove:
\begin{equation}
\exp ( -\sum_i ( \ln A_{0 ii} - t \frac{B_{ii}}{A_{0 ii}} ))
\end{equation}  

 We will use this result now in  the derivation of the renormalization group 
equation of the backscattering amplitude $t_m$.  

 Before giving the explicit calculation, let us sketch
 the RG analysis of the back scattering  amplitude 
which we want to perform in order to 
 find its effective dependence on the relevant energy scale
of the system $ E = max \{ T,  v/R \} $, temperature or
 level spacing.
 To this end we go to Matsubara frequency representation
 and integrate out all Fourier components of all the fields
in the functional integral in an infinitesimal energy window
$\Lambda/b < \omega_s < \Lambda $ where $ b = 1 + dl$.
 After that, we have to rescale the size of the
system by transforming $ \beta \rightarrow b \beta$ 
which also transforms the Matsubara frequencies as $\omega_l \rightarrow 
 \omega_l/b$.
 Then, one has to
 rescale the remaining  field variables such that 
 the effective back scattering  term in the action
has again the original form before integration over the fields in the energy
window, but now with a renormalizad back scattering amplitude.
  Then, we repeat this renormalization step M times
 until we reach the physically relevant
 energy scale E of the system, so that M is given 
 by $\Lambda/b^M = E$.

The  partition function 
of the FQH- annulus with a weak impurity at position $x=0$ 
 is given as a functional integral
in the momentum and Matsubara frequency representation:
\begin{eqnarray}
Z &=& \int \prod_{\omega_s } d J( \omega_s ) d  \phi_J ( \omega_s )
\nonumber \\ &&
\prod_{ n \neq 0}  d \phi_{-, n}(\omega_s ) d \phi_{+,
  n}(\omega_s ) ) \exp(-S)
\end{eqnarray}
   where $\omega_s = 2 \pi T s $ are the bosonic Matsubara frequencies.
 The action S is for the FQH- annulus with a weak impurity given by
\begin{equation}
S= S_0 + S_{BS},
\end{equation}
where $S_0$ is the action without back scattering between the edges,
\begin{eqnarray}
S_0 &=& T \sum_{\omega_s}
( - \frac{1}{2} \omega_s  \phi_J(\omega_s ) J ( - \omega_s )
 +  \frac{\nu v}{4 R} J(\omega_s) J(-\omega_s)  \nonumber \\
&+& \sum_{n>0} \sum_{p = \pm} (- \frac{n}{\nu} i 
\omega_s +  \frac{v_p}{\nu R_p} n^2 ) \phi_{p,
  n}(\omega_s ) \phi_{p, -n} ( -\omega_s) )
\nonumber \\
&+&  \frac{\nu v}{R}( - \varphi J(\omega_s=0) + \beta \varphi^2). 
\end{eqnarray}
where $v/R = \sum_p v_p/R_p/2$.
We consider the part of the action 
which describes back scattering by taking into account only 
Gaussian fluctuations about its minima,
that's when cos gives -1.
 Thus, we expand, $ \cos x = -1 + 1/2 ( x - (2 r + 1) \pi )^2 $,
and include the sum over the different minima r, in the partition function.
This Gaussian approximation of the back scatttering action thus gives in
Matsubara frequency representation a term
\begin{eqnarray}
S_{BS} &=&   \sum_{m=1}^{\infty} t_m ( - \beta( 1 - \frac{1}{2} 
( 2 r + 1 )^2 \pi^2 ) \nonumber \\
&-& ( 2 r +1 ) \pi m (  \phi_J( \omega_s =0 ) 
 +  
\sum_{n \neq 0} \sum_{ p = \pm} p \phi_p ( \omega_s = 0)  )
\nonumber
\\ &+& \frac{1}{2} m^2 
T \sum_{\omega_s} (  \phi_J ( \omega_s ) + 
\sum_{n}  \sum_{p = \pm} p \phi_{p,n} (\omega_s) )
\nonumber \\ && 
( \phi_J ( - \omega_s ) +
\sum_{n'} 
\sum_{p =\pm} p \phi_{p,n'} ( -\omega_s))).
\end{eqnarray}
 Note that while $S_0$ is diagonal both in Matsubara frequency and momentum
 space,
the back scattering term $S_{BS}$ is nondiagonal in momentum space.
However, as we have seen above,
 only the diagonal matrix elements of $S_{BS}$ contribute to
 the renormalization of the back scattering
 amplitudes
 $t_m$, so that we can easily proceed in the renormalization procedure
 by integrating out those fileds 
$J( \omega_s ), \phi_J ( \omega_s )$ 
$ \phi_{p, n} ( \omega_s ) $
 with energy in the interval $ \Lambda/b < \mid \omega_s \mid < \Lambda $.
As a result, we obtain 
to first order in $t_m$ the following additional terms:
\begin{equation}     
\sum_{\Lambda/b < \mid \omega_s \mid < \Lambda } 
 \sum_{m=1}^{\infty} \frac{1}{2} t_m m^2( \sum_p \sum_{n \neq 0} \frac{1}
{ A_{p,n}( \omega_s)}
 + \frac{1}{A_0( \omega_s )}),   
\end{equation}
where $ A_{p,n}( \omega_s) = n/\nu ( -i \omega_s + 2\pi v/L n ) $
 and $ A_0(\omega_s) = - R/( 2 \nu v) \omega_s^2 $.
Performing the summations over momentum and Matsubara frequency, we obtain,
noting that
 $ \Lambda \gg v/L $ that the  back scattering amplitudes $t_m$  become 
renormalized by 
\begin{equation}
 \beta \tilde{t}_m  = \beta  t_m ( 1- \nu m^2 \ln b
- \frac{\nu m^2 v}{2 \pi R \Lambda} (b-1) ) 
\end{equation}
 Before repeating this procedure of integrating over an infinitesimal
 energy window, we have to 
bring the action to the original form it had before the integration.  
To this end we have to blow the shrinked  size of the system up by $ \beta
 \rightarrow b \beta$ and accordingly substitute for the Matsubara 
 frequencies $ \omega_s \rightarrow \omega_s/b $.
Finally one has to transform the fields $ \phi_{p,n}( \omega_s ) \rightarrow C_1
 \phi_{p,n} ( \omega_s )  $,
 where the factor $C_1 = \sqrt{ \sum_{m=1}^{\infty} t_m  }/ \sqrt{
 \sum_{m=1}^{\infty} \tilde{t}_m } $ 
ensures that the back scattering action is transformed to the form it had 
 before the integration, in Gaussian approximation.
 Now we  start again by integrating
 over the next infinitesimal energy window.
We have  to keep on repeating this renormalization procedure, until
 the upper cutoff of the action is equal to a typical energy  E of the
 system: $ \Lambda/b^M = E$. Here, M is the number of renormalization cycles.
Thus, we obtain that the  back scattering  amplitude flows as a function of
 the typical energy E like
\begin{eqnarray}
t_m &\rightarrow &t_m b^{M-1} 
\prod_{j=1}^M  ( 
( 1 - m^2 \nu \ln b 
- \frac{\nu m^2 v}{L } \frac{b^{j-1}}{\Lambda} (b-1) 
) \nonumber \\ 
&\approx& t_m b^{M( 1-m^2
  \nu)} e^{- \nu m^2  v/( L \Lambda) b^M}. 
\end{eqnarray}   
which gives with $ b^M = \Lambda/E $:
\begin{equation}\label{teff}
t_m \rightarrow t_m ( E/\Lambda )^{m^2 \nu -1}e^{ - \nu m^2 v/L/E}.
\end{equation} 
 Thus, for the system with a finite level spacing $v/R$, 
 the relevant power law renormalization found in Ref.\cite{1},
 is modified  at temperatures close to the level spacing
 by a factor which depresses the tunneling amplitude.
 This behavior is well known as the Coulomb blockade effect,
 where here the charging energy is the level spacing $v/R$,
 the energy of an additional  particle on
 an edge.
 As the temperature is lowered below the level spacing, however,
 the typical energy scale E is the level spacing $v/R$,
 and the renormalized tunneling amplitude converges to
 a temperature independent value.

\section{Series and Integrals}
\label{a4}

 Here, we perform the summation over $n$ in the expression
\begin{equation}
D( \tau ) + D( -\tau) =  \nu \sum_p \sum_{n > 0}  \frac{1}{n}
\frac{\cosh ( n \pi v_p/(L_pT ) (1 - \mid \tau \mid ))}
{\sinh ( n \pi v_p/(L_pT) )}   ).   
\end{equation}
 We assume here 
that the level spacings on each edge are of similar magnitude.
 Then, we can treat two different temperature regimes:
 
 For  small temperatures below the level spacing $v/R$,
$ v/(R T ) \gg 1$, 
one can approximate
$\cosh(n \pi v/( T L) (1 -T \mid \tau \mid ) )/
 \sinh ( n \pi v/( L) ) \sim \exp( -n \pi v/L  \mid \tau \mid )$,
 and the summation over $n$ can be performed, yielding:
\begin{equation} \label{tsmall}
 D(\tau) + D(-\tau) = - 2 \nu \ln ( 1 - e^{- \pi v/L \mid \tau \mid } ).
\end{equation}  

 Denoting the ultraviolet cutoff by $\Lambda$ one obtains then also
\begin{equation} \label{tsmall0}
D(0) = -  \nu \ln ( \frac{v}{2 R \Lambda} ),
\end{equation}

 In the other limit, when the temperature exceeds  the level spacing, 
$v/(R T) \ll 1$,
 the summation can be transformed into a continous integral 
over $ x = v n/T/L $ 
with a lower
 cutoff $\epsilon \ll 1$ and one obtains:
 \begin{equation} 
D(\tau) + D(- \tau) = 2 \nu \int_0^{\infty} d x \frac{x}{x^2 + \epsilon^2}
\frac{\cosh ( \pi  ( 1 - T \mid \tau \mid ) x )}{\sinh ( \pi x ) }.
\end{equation}
 The integral can be performed ( Gradstein/ Ryshik 4.115.11., and
1.441.4. 
 \cite{grad})
and we obtain with $\epsilon \approx v/T/L$:
\begin{eqnarray} \label{tlarge}
D( \tau) + D( - \tau ) =  \nu T L/v  - 2 \nu  \ln [2 \sin ( \pi/2  T
   \mid \tau \mid ) ].
\end{eqnarray}
and with the ultraviolet cutoff $\Lambda$,
\begin{equation} \label{tlarge0}
D(0) = \nu/2 T L/v -  \nu \ln [   \pi T/\Lambda ].
\end{equation}

\newpage

{\bf Figures }

 Fig.2 :  \hspace{.5cm} The persistent current in a FQH- annulus with weak constriction,
at temperature to level spacing ratios, $t/(v/R) = .1,.05,.01$. 

 Fig. 3 : \hspace{.5cm}
 The persistent current in a clean FQH- annulus
at $T/(v/R) =.1,.05,.01$.

\begin{references}
\bibitem[*]{here} Present address,\\ E-mail: ketteman@thor.mpi-stuttgart.mpg.de,
\bibitem{laughlin83}  R.B. Laughlin, Phys. Rev. Lett. {\bf 50}, 1395(1983), 
\bibitem{5} D. J. Thouless, Phys. Rev. B {\bf 40}, 12034(1989), D. J. Thouless and Y. Gefen, Phys. Rev. Lett. {\bf 66} , 806(1991),
 Y. Gefen and D.J. Thouless   Phys. Rev. B {\bf 47}, 10423(1993
),
\bibitem{bye} N. Byers and C. N. Young, Phys. Rev. Lett. {\bf 7},46(1961).
\bibitem{prl} Y. Gefen, S. Kettemann, unpublished(1996),
\bibitem{2} X. G. Wen, Phys. Rev. B {\bf 43}, 11025(1991), 
\bibitem{1} C. L. Kane, M. P. A. Fisher, Phys. Rev. B {\bf 46}, 15233(1992),
\bibitem{3} C. de C. Chamon and X. G. Wen, Phys. Rev. Lett. {\bf 70},
2605(1993),
\bibitem{moon} K. Moon, H. Yi, C. L. Kane, S. M. Girvin, and Matthew P. A. 
Fisher  
, Phys. Rev. Lett. {\bf 71}, 4381(1993),
\bibitem{wen} X. G. Wen, Int. J. Mod. Phys. B {\bf 6}, 1711(1992),
\bibitem{halperin} B. I. Halperin, Phys. Rev. B {\bf 25}, 2185(1982), 
\bibitem{exact} P. Fendley, A. W. W. Ludwig, and H. Saleur, Phys. Rev. Lett.
 {\bf 75}, 2196 (1995), and U. Weiss, preprint (1996),   
\bibitem{avishai} Y. Avishai, M. Kohmoto, Phys. Rev. Lett. {\bf 71}, 279(1993), 
\bibitem{kim} S. Lee, H. C. Kwon, K. W. Park, M. Shin, J. S. Yuk, S. Kim,
  E. Lee, preprint ( 1996),
\bibitem{dror} D. Orgad, S. Levit, Phys. Rev. B {\bf 53}, 7964(1996),
\bibitem{oreg} Y. Oreg and A. M. Finkel'stein, Phys. Rev. Lett.
 {\bf 74}, 3668(1995),
\bibitem{haldane} F. D. M. Haldane, J. Phys. C {\bf 14}, 2585(1981),
\bibitem{stone} M. Stone and M. P. A. Fisher, ``Luttinger states at the 
edge'', NSF-ITP-94-15 
\bibitem{negele} John W. Negele, Henri Orland, {\it Quantum Many- Particle
 Systems}, Addison- Wesley (1988).  
\bibitem{grad} Gradstein, Ryshik, Verlag Harri Deutsch, Frankfurt/M.(1981)
\end{references}
\end{document}